\documentclass{elsarticle}

\usepackage{graphicx}
\usepackage{amssymb}
\usepackage{amsthm}
\usepackage{amsmath}
\usepackage[utf8]{inputenc}
\usepackage{algorithm}
\usepackage{algpseudocode}
\usepackage{xcolor}

\usepackage{lineno}
\usepackage{url}

\newcommand{\tuple}[1]{\ensuremath{\langle #1} \rangle}

\newtheorem{definition}{Definition}

\begin{document}

\begin{frontmatter}

\title{Detecting coherent explorations in SQL workloads}

\cortext[mycorrespondingauthor]{Corresponding author}

\author[ut]{Ver\'{o}nika Peralta\corref{mycorrespondingauthor}}
\ead{veronika.peralta@univ-tours.fr}

\author[ut]{Patrick Marcel}
\ead{patrick.marcel@univ-tours.fr}

\author[ut]{Willeme Verdeaux}
\ead{willeme.verdeaux@etu.univ-tours.fr}

\author[ut]{Aboubakar Sidikhy Diakhaby}
\ead{aboubakar-sidikhy.diakhaby@etu.univ-tours.fr}

\address[ut]{University of Tours, Tours, France}

\begin{abstract}
This  paper  presents a proposal 
aiming at better understanding 
a 
workload of 
SQL queries and detecting coherent 
explorations hidden within the workload.
In particular, our work investigates
SQLShare \cite{DBLP:conf/sigmod/JainMHHL16}, a database-as-a-service platform targeting scientists and data scientists with minimal database experience,
whose workload was made available to the research community.
According to the authors of \cite{DBLP:conf/sigmod/JainMHHL16},
this workload is the only one containing primarily ad-hoc
hand-written queries over user-uploaded datasets.
We analyzed this workload by extracting features that
characterize SQL queries and we show how to use these features to separate sequences of SQL queries  
into meaningful explorations. 
We ran several tests 
over various query workloads
to validate empirically our approach. 
\end{abstract}

\end{frontmatter}

\section{Introduction}
\label{sec:introduction}

Analyzing a database workload 
offers many practical interests, from
the monitoring of database physical access structures
\cite{DBLP:conf/vldb/ChaudhuriN07}
to the generation of user-tailored 
collaborative query recommendations 
for interactive exploration \cite{DBLP:journals/tkde/EirinakiAPS14}.
There has been much attention lately
devoted to the analysis of user past activities
to support Interactive Database Exploration (IDE)
\cite{DBLP:conf/sigmod/IdreosPC15}.
OLAP analysis of data cubes is a particular case of IDE, that takes advantage of simple primitives like drill-down or slice-and-dice for the navigation of
multidimensional data.
These particularities enable the design
of approaches for characterizing user explorations
in how focus they are
\cite{DBLP:conf/adbis/DjedainiLMP17},
in how contributive a query is to the exploration
\cite{DBLP:journals/is/DjedainiDLMPV19}, or even in how
to ensure that a sequence of analytical queries forms a
coherent exploration
\cite{DBLP:conf/dawak/RomeroMAPB11}.

Transposing these works to regular, non multidimensional
SQL workloads raises many challenges.
Even if a sequence of SQL queries is issued
to explore the database instance,
non multidimensional relational schemas 
do not have regularities one expects from
the multidimensional model, 
explorations may not be expressed
through roll-up or drill-down operations,
SQL queries may deviate from the traditional
star-join pattern commonly used for analytical purpose, 
etc.
    
In this paper, we present an 
approach for analyzing SQL workloads,
concentrating on the SQLShare workload
of hand-written\footnote{Consistently with 
the authors of \cite{DBLP:conf/sigmod/JainMHHL16},
we use the term
hand-written to mean, in this context, that the query is introduced 
 manually  by a human user, which reflects genuine interactive
human activity over a dataset, with consideration between
two consecutive queries.} queries over
user-uploaded datasets.
This workload includes raw sequences of queries
made by some users, without further information
on their intention.
One of our objectives is to investigate
whether this workload contains actual
exploration activities, and more particularly
how to extract such explorations.
In what follows, we consider that
an \textit{exploration} is a coherent sequence of hand-written  queries, that all share the same goal of fulfilling a user's information need that may not be well defined initially, while a \textit{session} is just a raw sequence of queries, possibly containing several explorations.  
Identifying such exploration activities has several applications in 
IDE, 
like  understanding users' information needs, identifying "struggling" during the exploration, providing better query recommendations, etc. This is important since, usually, systems used for IDE do not offer such facilities. 

To identify explorations from a SQL workload, 
we use a technique first proposed in \cite{DBLP:journals/is/DjedainiDLMPV19}
to score the quality of OLAP explorations.
This technique  consists of characterizing a query
by a set of simple
features that are intrinsic to a query or
that relate the query to its neighbor in the
sequence. 
While in \cite{DBLP:journals/is/DjedainiDLMPV19} this technique
of feature extraction was used with supervised
machine learning to score the quality of OLAP explorations, 
in the present work we use these features to
partition an SQL workload into coherent
explorations, 
investigating three different alternatives for session segmentation: 
\begin{itemize}
    \item unsupervised learning: our first method is based only on similarity between contiguous queries, 
    \item supervised learning: our second method uses transfer learning to reuse a model trained over a dataset where ground truth is available,  
    \item weak supervision: our third method uses weak labelling to predict the most probable segmentation from  heuristics meant to label a training set.
\end{itemize}


The work we present here  is a follow-up to \cite{DBLP:conf/dolap/PeraltaVRM19}, where the first method
was originally motivated and introduced. Here we have improved our previous
work under several substantial aspects: (i) the adding of new query features and the study of feature correlations; (ii) the proposal of two additional methods for session segmentation (in \cite{DBLP:conf/dolap/PeraltaVRM19} only similarity-based segmentation was considered), and (iii) the experimental evaluation of the proposed methods, including a comparative analysis. 


The paper is organized as follows.
The next section discusses related work.
Section \ref{sec:preliminaries} presents 
our model of queries,
tailored for  SQL queries.
Section \ref{sec:extraction} 
details the features considered and how they are extracted. 
Section \ref{sec:sessions} introduces our segmentation strategies
and 
Section \ref{sec:tests} reports the results of the tests we conducted.
Section \ref{sec:conclusion}
concludes and draws perspectives.
\section{Related Work}
\label{sec:related}

In this section we present related work concerning real SQL workloads and workload analysis.

\subsection{Real SQL workloads} 

\paragraph{SQLShare}
The SQLShare workload is the result of a multi-year SQL-as-a-Service experiment \cite{DBLP:conf/sigmod/JainMHHL16}, allowing any user with minimal database experience to upload their datasets on-line and manipulate them via SQL queries. What the authors wanted to prove with this experiment is that SQL is beneficial for data scientists. They observed that most of the time people use scripts to modify or visualize their datasets instead of using the SQL paradigm. Indeed, most user needs may be satisfied by first-order queries, that are generaly much simpler than a script, but have the initial cost of creating a schema, importing the data and so on. SQL-as-a-Service frees the user of all this prior work with a relaxed SQL version. 

The SQLShare workload is composed of 11,137 SQL statements, 57 users and 3,336 user's datasets. 
To the best of our knowledge, as reported by  the authors of \cite{DBLP:conf/sigmod/JainMHHL16}, this workload is the only one containing primarily ad-hoc hand-written queries over user-uploaded datasets.
As indicated in the introduction,  hand-written means that the query is introduced 
 manually  by a human user, which reflects genuine interactive
human activity over a dataset, with consideration between
two consecutive queries.

The SQLShare workload is analyzed in 
\cite{DBLP:conf/sigmod/JainMHHL16}, particularly
to verify the following assumption:
\begin{quote}
    "We hypothesized that SQLShare users would write queries that are more complex individually and more diverse as a set, making the corpus more useful for designing new systems."
\end{quote}

The authors indeed showed empirically that the queries 
in the SQLShare workload are complex and diverse.
They also analyzed the churn rate of SQLShare users
and conclude that most users exhibit a behavior
that suggest an exploratory workload.
However, there is not ground truth, nor preliminary guidelines, about how to detect coherent explorations within the workload. Timestamps are neither included.
To our knowledge, and again as reported by the authors
of \cite{DBLP:conf/sigmod/JainMHHL16}, this workload is 
one of the two workloads publicly available to the
research community, the other being the SDSS workload.


\paragraph{SDSS workload}

SkyServer is an Internet portal to the Sloan Digital Sky Survey Catalog Archive Server; its Web and SQL logs are public \cite{skyserver-traffic-report-the-first-five-years}.
The SQL log was produced by a live SQL database supporting  
both ad hoc hand-authored queries as well as queries generated from a point-and-click GUI. 
Many queries in the SDSS are actually not hand-written; they were generated by applications such as the Google Earth plugin or the query composer from the SkyServer website.
Their cleaning and normalization took several months effort.

Sessions in this log were detected using heuristics:
\begin{quote}
     "We arbitrarily start a new session when the previous page view from that IP address is more than 30 minutes old, i.e., a think-time larger than 30 minutes starts a new session. [...]  Wong and Singh \cite{DBLP:conf/mmm/BhattaraiWS07} chose the same 30 minute cutoff and we are told that MSN and Google use a similar heuristic."
\end{quote}

The authors of \cite{skyserver-traffic-report-the-first-five-years} also
acknowledge the difficulty of  extracting
human sessions from all those collected:
\begin{quote}
    "We failed to find clear ways to segment user populations.
    We were able to ignore the traffic that was administrative or was eye-candy, leaving us with a set of 65M page views and 16M SQL queries. We organized these requests into about 3M sessions, about half of which were from spiders. The residue of 1.5M sessions had 51M page views and 16M SQL queries – still a very substantial corpus.
    [...]
    Interactive human users were 51\% of the sessions, 41\% of the Web traffic and 10\% of the SQL traffic. 
    We cannot be sure of those numbers because we did not find a very reliable way of classifying bots vs mortals."
\end{quote}

Bots are programs that automatically crawled the SDSS
and launch SQL queries. Such traffic cannot be classified
as proper interactive data exploration with human consideration.

In \cite{DBLP:conf/sigmod/JainMHHL16}, the authors
compared the SQLShare workload and that of the SDSS, and
conclude:
\begin{quote}
"SQLShare queries on average tend to be more complex and more diverse than those of a conventional database workload generated from a comparable science domain: the Sloan Digital Sky Survey (SDSS)."
\end{quote}

\paragraph{Smaller SQL datasets} 

We are aware of other available SQL workloads.
For instance, Kul et al. 
\cite{DBLP:journals/tkde/KulLXCKU18}
analyze three specific query sets.
The first one, Student assignments gathered
by IIT Bombay, is made of a few hundreds queries
answering homework assignments.
The second dataset, publicly available, consists 
of around 200 queries gathered over 2 years from 
student exams at University of Buffalo.
Actually, it does not contain real explorations but sets of queries supposed to be similar, and experiments in a preliminary study \cite{DBLP:conf/dolap/PeraltaVRM19} show that it is not appropriate for testing session segmentation.
The third dataset consists of SQL logs that capture all
database activities of 11 Android phones for a period of
one month. The log consists of 1,352,202 SELECT statements that, being generated by an application,
correspond to only 135 distinct query strings.


\subsection{Other database workloads with ground truth}

In addition to SQL workloads, we present two additional workloads consisting of 
logs of multidimensional queries, devised by real users, that we have used in our previous works \cite{DBLP:journals/is/DjedainiDLMPV19}.
These workloads, while containing a particular case of queries (star-join queries), are interesting because a ground truth (the set of queries corresponding to each user exploration) is available, allowing the evaluation of our approach.

\paragraph{Open dataset}
The first dataset, named \textit{Open} in what follows, consists of navigation traces collected 
in the context of a French project on energy vulnerability. 
These traces were produced by 8 volunteer students 
of a Master's degree in Business Intelligence,  
answering  fuzzy information needs defined by their lecturer,
to develop explorative OLAP navigations using Saiku\footnote{http://meteorite.bi/products/saiku} over 
three cubes instances.
The main cube is organized as a star schema with 19 dimensions, 68 (non-top) levels, 24 measures, and contains 37,149 facts recorded in the fact table.
The other cubes are organized in a similar way.
From this experiment, we reuse $16$ sessions, representing $28$ explorations and $941$ queries.
The ground truth is a manual segmentation
made by the lecturer based on some guessed notion of user goal and supported by timestamps. 
Notably, automatic segmentation was not the purpose of the work at the time 
manual segmentation was done.

\paragraph{Enterprise dataset}
The second dataset, named \textit{Enterprise}, consists of navigation traces of 14 volunteers among SAP employees, in the context of a previous study on discovering user intents  \cite{DrushkuLMP19}.
We set 10 business needs, and volunteers were asked to analyze some of the 7 available data sources to answer each of the 10 business needs, 
using a SAP prototype that supports keyword-based BI queries\footnote{Patent Reference: 14/856,984 : BI Query and Answering using full text search and keyword semantics}. 
In total, this dataset contains $24$ sessions, corresponding to $104$ user explorations and accounting for $525$ queries.
Volunteers were explicitly requested to express to what information needs they were answering,  which constitutes our ground truth for this dataset. \\


Notably, in Open and Enterprise datasets, users did not have to write any SQL code,
contrarily to SQLShare.  
Indeed, Saiku and the SAP prototype generated queries from users high-level operations. However, in both cases, users devised real explorations, taking the time to analyse results before devising new querys.
Users of the Open dataset were Master students learning data analysis skills, users of the Enterprise dataset were developers with varied analysis skills, while SQLShare users are anonymous end-users and there is no knowledge about their analysis skills.


\subsection{Workload analysis}


Other scientific domains
close to Database, like Information Retrieval or Web Search,
have a long tradition of log analysis
aiming at facilitating the searcher's task \cite{DBLP:books/cu/W2016}.
Many works extract features from queries or search sessions
and use them to disambiguate the session's goal, 
to generate recommendations, 
to detect struggling in sessions, etc.
Since databases tend to be more used in an exploratory or analysis fashion,
as evidenced by the SQLShare workload,
it is not a surprise that many recent works 
pay attention to the analysis of database workloads,
in addition to those works analyzing workload
for optimization or self-tuning purposes.
We present some recent advances in this area,
differentiating by the type of logs 
(OLAP logs and 
SQL logs
).

\paragraph{Analyzing and detecting OLAP explorations}

Logs of OLAP analyses  are simpler than SQL ones in the sense
that they feature multidimensional queries that can
easily be interpreted in terms of OLAP primitives
(roll-up, drill-down, slice-and-dice, etc.).
In one of our previous works \cite{DBLP:conf/dawak/RomeroMAPB11},
we proposed an approach for detecting OLAP
analyses phrased in SQL, by converting SQL queries
into OLAP queries and then checking if two consecutive
queries are sufficiently close in terms of OLAP operations.
In our more recent work, we used supervised learning
to identify a set of query features allowing to characterize
focus zones in OLAP explorations \cite{DBLP:conf/adbis/DjedainiLMP17},
or to identify queries that better contribute to an exploration  
\cite{DBLP:journals/is/DjedainiDLMPV19}.
The present work can be seen as a continuation of those previous 
works, since we have the same objective as  \cite{DBLP:conf/dawak/RomeroMAPB11}
and use the same technique as \cite{DBLP:journals/is/DjedainiDLMPV19}.
The main differences with these previous works is that we make no assumption about the type of queries in the workload (particularly, they may not be multidimensional queries), and we have no ground truth (i.e., no human manual inspection of each query) on the workload.

\paragraph{Analyzing SQL logs}

SQL workload analysis has recently attracted 
attention beyond query optimization,
for instance for query recommendation \cite{DBLP:journals/tkde/EirinakiAPS14},
 query autocompletion \cite{DBLP:journals/pvldb/KhoussainovaKBS11},
or user interest discovery \cite{DBLP:conf/edbt/NguyenBBGHM15}.
%
All these works use the SDSS workload for their tests.
Embedded SQL code is  analyzed in \cite{DBLP:conf/scam/BrinkLV07}
to measure its quality, mainly for maintainability purpose. The authors quantify
quality based on the number of operators (joins, unions), operands (tables, subqueries) and variables in the SQL code,
experimenting with SQL codes embedded in PL/SQL, COBOL and Visual Basic.
 Jain et al. ran a number of
tests on the SQLShare workload \cite{DBLP:conf/sigmod/JainMHHL16}, 
some of them being reported above,
showing the diversity and complexity of the workload.
In \cite{jainComplexity}, Vashistha and Jain analyze  the complexity of queries in the SQLShare workload, in terms of the following query features:  number of tables, number of columns, query length in characters, numbers of  operators (Scan, Join, Filter),  number of comparison operators (LE, LIKE, GT, OR,
AND, Count), and the query run-time.
They define two complexity metrics from these features: the Halstead measure (traditionally used to measure programs complexity) and a linear combination whose weights are learned using regression.
Finally, a recent work investigated various
similarity metrics over SQL queries, aiming
at clustering queries \cite{DBLP:journals/tkde/KulLXCKU18}
for better workload understanding.
The authors  run their tests on smaller SQL sets,
as indicated above.



There is very few works about segmentation of SQL queries. Several works (e.g. \cite{skyserver-traffic-report-the-first-five-years}), needing to detect sessions as part of their preprocessing tasks, implemented simple heuristics, inspired from detection of sessions in Web logs (e.g. \cite{Wong06characterizationand}). Basically, a new session starts after 30 minutes of user inactivity.
%
%
A different approach, based on supervised learning, is used by Khoussainova et al. \cite{DBLP:journals/pvldb/KhoussainovaKBS11}, with the goal of reducing the size of the query log 
for improving query autocompletion. 
Authors claim that  
users do many tries until obtaining the ``good" query and
higher quality queries appear at the end of a segment of similar queries.
They propose a Query Eliminator module that segment the query log, stitch similar segments and drop many queries.  
They use a perceptron-based classifier to decide whether two consecutive queries belong to the same segment. Features used for segmentation include time interval between queries, cosine similarity between query clauses and relationship among 
abstract syntax trees of the queries. 
Their approach is tested on manually annotated logs and compared to the timestamp-based heuristic used for Web logs.
Nevertheless, the perceptron is not trained to detect explorations, but segments of similar queries, in order to reduce the log size. 
%

To our knowledge, our work is the first devoted to segment 
hand-written SQL queries into meaningful explorations, 
without ground truth or timestamps, and without making assumptions
about the position of quality queries in a session.
%




\section{Preliminaries}
\label{sec:preliminaries}

This section introduces the SQLShare workload and describes our hypothesis and preprocessing.

\subsection{SQLShare workload preprocessing}

From the 11,137 SQL statements we kept 10,668 corresponding to SELECT statements. The remaining statements (mainly updates, inserts and deletes) were filtered out. 

We implemented a preliminary 
segmentation following a simple heuristic: keeping together the sequences of consecutive queries of a same user. 
As a result of the initial segmentation we obtained 451 sessions, counting between 1 and 937 queries (average of 23.65 queries per session, standard deviation of 75.05 queries).
Furthermore, we made the initial hypothesis that queries appear in chronological order in the SQLShare workload.
We noted that the queries of the workload do not
come with timestamps, and we contacted 
the authors of the original SQLShare paper
\cite{DBLP:conf/sigmod/JainMHHL16} who 
confirmed 
that the query order in the workload
may not reflect the order in which queries
were launched.
Therefore, the disparate distribution of queries along sessions, in addition to some extremely long sessions, possibly disordered, calls for a smarter way of segmenting sessions. Furthermore, segmentation strategy should be based only on query text, as for privacy issues, only a portion of instances is available.

\subsection{Query and session abstractions}
 

In what follows, we use the term \emph{query} to denote the text of an SQL SELECT statement.  
We represent a query as a 
collection of fragments extracted from the query text, namely, projections, selections, aggregations, 
tables and attributes.
The first four 
fragments abstract the most descriptive parts of a SQL query, and are the most used in the literature (see e.g., \cite{DBLP:journals/pvldb/KhoussainovaKBS11,DBLP:journals/tkde/EirinakiAPS14}). 
The set of attributes (i.e., all attributes appearing somewhere in the query text) is included as it was used in 
preliminary studies of SQLShare complexity \cite{DBLP:conf/sigmod/JainMHHL16, jainComplexity}  
and intuitively, it may complement the other query fragments, specially for those attributes appearing in complex SQL clauses. 
Note that we do not restrict to SPJ (selection-projection-join) queries. Indeed, we consider all queries in the SQLShare workload, some of them containing arbitrarily complex chains of sub-queries.

\begin{definition}[Query] 
\label{def:query}
A {\em query} over database schema $DB$ is a quintuple 
of query fragments
$q= \tuple{P,S,A,T,At}$ 
where:
\begin{enumerate}
\item $P$ is a set of expressions (attributes or calculated expressions) appearing in the main SELECT clause (i.e. the outermost projection). We deal with * wild card by replacing it by the list of attributes it references.
\item $S$ is a set of atomic Boolean predicates, 
whose combination (conjunction, disjunction, etc.) defines the WHERE and HAVING clauses appearing in the query. We considered indistinctly all predicates appearing in the outermost statements as well as in inner sub-queries.
\item $A$ is a set of aggregation expressions  appearing in the query text. We considered indistinctly all expressions appearing in the outermost statements as well as in inner sub-queries, disregarding the SQL clause where they appear. 
\item $T$ is a set of tables appearing in FROM  clauses (outermost statement and inner sub-queries). Views, sub-queries and other expressions appearing in FROM clauses are parsed in order to obtain the referenced tables.  
\item $At$ is a set of attributes appearing explicitly in the query (outermost statement and inner sub-queries). Expressions, views, sub-queries and other clauses are parsed in order to obtain the referenced attributes. This allows to consider all attributes, even those that are part of atypical or less-frequently-used clauses. 
However, the * wild card is not replaced with the list of referenced attributes, as for projections. 

\end{enumerate}
\end{definition}

Note that although we consider tables, 
selections and aggregations occurring in inner sub-queries, 
we limit to the outermost queries for projections
, as they correspond to attributes actually visualized by the user. 
We intentionally remain independent of presentation and optimization aspects, specially the order in which attributes are projected (and visualized by the user), the order in which tables are joined, etc.

Finally, a \emph{session} is a sequence of queries of a user over a given database.

\begin{definition}[Session]
Let $DB$ be a database schema.
A session $s=\tuple{q_{1}, \ldots, q_{p}}$
over $DB$ is a sequence of queries over $DB$.
We note $q \in s$ if a query $q$ appears in the session $s$, and  $session(q)$ to refer to the session where $q$ appears.  
\end{definition}

\section{Feature extraction}
\label{sec:extraction}

In this section, we define a set of features to quantitatively describe different aspects of a SQL query and its context.
We then describe the extraction procedure and the obtained scores.

\subsection{Feature description}
\label{sec:metrics}

%
For each query, we extract a set of simple features computed from the query text and its relationship with other queries in 
a session.
%
The set of features is inspired from our previous work 
\cite{DBLP:conf/adbis/DjedainiLMP17, DBLP:journals/is/DjedainiDLMPV19}, which models OLAP queries as a set of features capturing typical OLAP navigation. 

We intend to cover various aspects of a query in order to support different types of analysis and modeling based on query features.
In particular, 8 features form the  core of our proposal; they count the number of projections, selections, aggregations and tables and the number of common projections, selections, aggregations and tables.
Other features combine them, providing useful computations (edit distance and Jaccard index) and 
other ones (e.g. number of attributes and number of characters) are informative and
allow the comparison with preliminary works on SQLShare complexity \cite{DBLP:conf/sigmod/JainMHHL16, jainComplexity}.

From now on, to remove any ambiguity, we use
the term metric to denote the functions that
score query properties. 
The term feature is reserved to denote the score
output by the function.

For the sake of presentation, we categorize metrics as follows:
i) intrinsic metrics, i.e., only related to the query itself,
and
ii) relative metrics, i.e., also related to the query's predecessor in the session. 
Table \ref{table:query_metrics} presents an overview of the metrics.



\begin{table}[!ht]
\begin{center} {
\begin{tabular}{l|l}

\hline
\multicolumn{2}{c}{Intrinsic metrics} \\
\hline
NoP			&  Number of projections (attributs and expressions)	\\
NoS 		&  Number of selections (filtering predicates) \\
NoA 		&  Number of aggregations 	\\
NoT			&  Number of tables	\\
NoAt		&  Number of attributes	\\
NoCh		&  Number of characters	\\
\hline
\multicolumn{2}{c}{Relative metrics} \\
\hline
NCP 		&  Number of common projections, with previous query \\
NCS 		&  Number of common selections, with previous query \\
NCA 		&  Number of common aggregations, with previous query \\
NCT 		&  Number of common tables, with previous query \\
RED 		&  Relative edit distance (effort to express a query starting  \\
	& from the previous one) \\
JI 		    &  Jaccard index of common query fragments, with previous \\
    & query  \\
\hline
\end{tabular}
}
\caption{metrics for SQL queries
\label{table:query_metrics} }
\end{center}
\end{table}

For all definitions given in this section, let 
$q_k = \tuple{P_k,S_k,A_k,T_k,At_k}$ 
be the query occurring at position $k$ in the session $s$ over the instance $I$ of schema $DB$. 
All the queries we considered are supposed to be well formed, and so we do not deal with query errors.

For the moment, we only considered metrics based on query text. Further metrics may be defined if the database instance is taken into account (for example, number of tuples in query result, precision and recall of query result w.r.t. previous query, execution time, etc.). But the computation of such metrics implies the execution of every query in the SQLShare dataset, which is not always available 
for confidentiality reasons (i.e. users did not agree to share their data),
and thus considerably reduces the set of queries. We left such studies to future work. 


\subsubsection{Intrinsic metrics}

Intrinsic metrics are those that can be computed only considering  the query $q_k$, independently of the session $s$ and other queries in $s$. In other words, these metrics will give the same score to $q_k$, independently of $s$.

\paragraph{Number of Projections}
$NoP(q_k)$ represents the number of projections (attributes and expressions) that are projected by the user. Expressions projected in inner sub-queries, but not projected by the outer query, are not considered as they do not appear in the query result. 
\begin{equation}
\begin{aligned}
NoP{}(q_k) = card(P_k)
\end{aligned}
\end{equation}

\paragraph{Number of Selections}
$NoS(q_k)$ represents the number of selections (elementary Boolean predicates) that appear in the query text, both in outer and inner sub-queries.  
\begin{equation}
\begin{aligned}
NoS{}(q_k) = card(S_k)
\end{aligned}
\end{equation}

\paragraph{Number of Aggregations}
$NoP(q_k)$ represents the number of aggregations 
that appear in the query text, both in outer and inner sub-queries.
\begin{equation}
\begin{aligned}
NoA{}(q_k) = card(A_k)
\end{aligned}
\end{equation}


\paragraph{Number of Tables}
$NoP(q_k)$ represents the number of tables appearing in query text, both considering outer and inner sub-queries.  
\begin{equation}
\begin{aligned}
NoT{}(q_k) = card(T_k)
\end{aligned}
\end{equation}

\paragraph{Number of Attributes}
$NoP(q_k)$ represents the number of attributes appearing in query text, both considering outer and inner sub-queries.  
\begin{equation}
\begin{aligned}
NoAt{}(q_k) = card(At_k)
\end{aligned}
\end{equation}

\paragraph{Number of Characters}
$NoP(q_k)$ represents the number of characters of the query text,  both considering outer and inner sub-queries.  
\begin{equation}
\begin{aligned}
NoCh{}(q_k) = length(q_k)
\end{aligned}
\end{equation}

\subsubsection{Relative metrics}

Relative metrics are those that are computed comparing the query $q_k$ to the previous query in the session $s$, 
$q_{k-1} = \tuple{P_{k-1},S_{k-1},A_{k-1},T_{k-1},At_{k-1}}$. 
For the first query of $s$, i.e. $q_1$, we consider as predecessor the "empty" query $q_0 = \tuple{\emptyset, \emptyset, \emptyset, \emptyset, \emptyset}$. 
All the following metrics are defined for $k\geq 1$.

\paragraph{Number of Common Projections}
$NCP(q_k,q_{k-1})$ counts the number of common projections of $q_k$ relatively to $q_{k-1}$.
\begin{equation}
\begin{aligned}
NCP(q_k, q_{k-1}) = 
card(P_k \cap P_{k-1})  
\end{aligned}
\end{equation}

\paragraph{Number of Common Selections}
$NCS(q_k,q_{k-1})$ counts the number of common selections of $q_k$ relatively to $q_{k-1}$.
\begin{equation}
\begin{aligned}
NCS(q_k, q_{k-1}) = 
card(S_k \cap S_{k-1})
\end{aligned}
\end{equation}

\paragraph{Number of Common Aggregations}
$NCA(q_k,q_{k-1})$ counts the number of common aggregations of $q_k$ relatively to $q_{k-1}$.
\begin{equation}
\begin{aligned}
NCA(q_k, q_{k-1}) = 
card(A_k \cap A_{k-1})
\end{aligned}
\end{equation}

\paragraph{Number of Common Tables}
$NCT(q_k,q_{k-1})$ counts the number of common tables of $q_k$ relatively to $q_{k-1}$.
\begin{equation}
\begin{aligned}
NCT(q_k, q_{k-1}) = 
card(T_k \cap T_{k-1})
\end{aligned}
\end{equation}


\paragraph{Relative Edit Distance}
$RED(q_k,q_{k-1})$ represents the edition effort, for a user, to express the current query starting from the previous one.
It is strongly related to query fragments, 
and computed as the minimum number of atomic operations between queries, by considering the operations of adding/removing a projection, selection, aggregation or 
table.  
The considered cost for each observed difference (adding/removing) is the same.
\begin{equation}
\begin{aligned}
RED(q_k,q_{k-1}) = card(P_k - P_{k-1}) + card(P_{k-1} - P_k) \\
+ card(S_k - S_{k-1}) + card(S_{k-1} - S_k) \\
+ card(A_k - A_{k-1}) + card(A_{k-1} - A_k) \\
+ card(T_k - T_{k-1}) + card(T_{k-1} - T_k) \\
\end{aligned}
\end{equation}

\paragraph{Jaccard Index}
$JI(q_k,q_{k-1})$ represents the ratio between the common query fragments 
(projections, selections, aggregations and 
tables) 
and the union of query fragments. 
\begin{equation}
\label{eq:JaccardIndex}
\begin{aligned}
JI(q_k,q_{k-1}) = \frac{card(Fragments(q_k) \cap Fragments(q_{k-1})}
{card(Fragments(q_k) \cup Fragments(q_{k-1})}
\end{aligned}
\end{equation}
where Fragments($q_k$) = $P_k \cup S_k \cup A_k \cup T_k$.





\subsection{Extraction protocol}
\label{sec:protocol}
This section briefly describes the procedure for extracting query features. We proceed in 3 steps: 
\begin{enumerate}
    \item Filter the SQL workload in order to keep only SELECT statements, i.e. discard updates, inserts, etc.
    \item Extract query fragments (sets of projections, selections, aggregations, 
    tables and attributes) and compute length, for each query. 
    \item Compute query features from query fragments.
\end{enumerate}
For the first and second step we developed a C\# script using the MSDN TSQL Parser. Removing all the non SELECT statements was  straightforward. However, extracting query fragments  required to deal with several particular cases. 
The major challenge was extracting projections.
First, for getting the projections visualized by the user, we need 
to detect the outermost SELECT clause of a query and then extract the SELECT elements.
When there is a * wild card 
in the query (e.g. as in \textit{SELECT * FROM T}), our script reaches the outermost FROM clause (looking for \textit{T}). When \textit{T} is a table, the script accesses schema metadata and obtains all the table attributes. 
If there is one or more sub-queries in the FROM clause, it repeats the previously described pattern until it finds either a table or a set of SELECT elements. Views and WITH clauses were treated in a similar way. For now, 
we do not take into account the queries having more than one '*' in their SELECT elements. (i.e : \textit{SELECT t1.*,t2.*,a,b FROM t1,t2}).

Note that this procedure for resolving * wild cards relies in the existence of schema metadata, i.e., having access to the corresponding datasets in order to obtain the list of attributes. 
However, some datasets are referenced in queries but are not present in the SQLShare released data because the user decided not to share them. For such queries (near 18\% of the query workload), we could not resolve the list of projections and we needed to estimate the number of projections during third step (computing features).

The aggregations were obtained with a Parser's function that detects all the function calls within a query. 
The selections are all the atomic Boolean expressions contained in the queries and their subqueries. 
The extraction of tables and attributes is straightforward using specific Parser's functions and regular expressions.
At this stage of our work we do not deal with 
predicate containment.

The third step
computes query features
as described in Equations 1 to \ref{eq:JaccardIndex}.  
For the unresolved * wild cards, 
we estimated both the number of projections and the number of common projections,
by taking advantage of the 
other queries in the exploration that list  attributes of those tables. 
We used linear regression to estimate the remaining cases, with the AUTO-SKLEARN  Python module \cite{NIPS2015_5872}, which is a module aiming at automatically choosing and parametrizing a machine learning algorithm for a given dataset, at a given cost (i.e., the time it takes to test different algorithms).


More precisely, we computed 
two distinct regressions, one for NoP and another one for  NCP. 
The methodology is the same for both and consists of the following: 
\begin{enumerate}
    \item Each query is represented by 23 features : NoS, NoA, NoT, NCS, NCA, NCT 
    (neither NoP nor NCP since they are the target of the regression), the min, max, average and standard deviation of the NoP, NoS, NoA, and NoT, grouped by the
    session 
    the query belongs to
    and the number of queries in the session. 
    \item The queries where NoP=0 (and consequently NCP=0) are removed from the data set. These queries are the ones where NoP and NCP have to be estimated.
    \item  The data set is then split in 80/20 for cross-validation, and the AUTO-SKLEARN regression mode is used to fit the best regression. For the first regression, NoP is the target, for the second, NCP is the target. 
    The maximum time to find a model is set to 180 seconds and the score used to measure the accuracy of the regression is $R^2$ (coefficient of determination) regression score function. 
    \item The $R^2$ score of each regression is used to predict NoP (respectively NCP) for the queries removed at step 2.
    \item Degenerated cases (e.g., a predicted number of common projection that is greater than the predicted number of projections) are fixed manually. 
\end{enumerate}







\subsection{Analysis of query features for the SQLShare dataset}
\label{sec:extr-sql}

\begin{figure}
\begin{center}
\includegraphics[width=12cm]{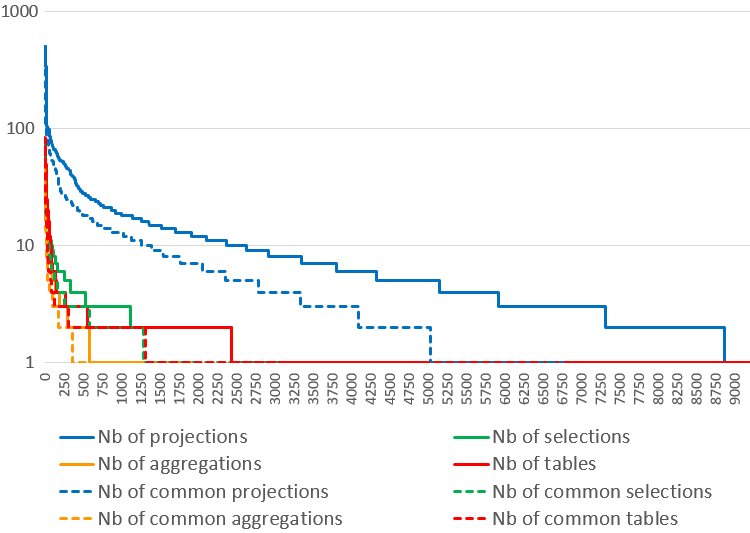}
\end{center}
\caption{Value distribution of main query features in the SQLShare dataset. 
}
\label{fig:query-features}
\end{figure}

Table \ref{table:trends-query} summarizes the results of feature extraction. Value distributions of the main features are shown in Figure \ref{fig:query-features}.
These results are slightly different from the ones presented in our preliminary study \cite{DBLP:conf/dolap/PeraltaVRM19} because 
we used strict sets instead of multisets for representing some query fragments (e.g. we count only 1 table in the following query: SELECT t1.a, t2.a FROM T t1, T t2).

\begin{table}[!ht]
\begin{center}
{\small{ 
\begin{tabular}{lrrrrrrrrr}

\textbf{Metric} & \textbf{Avg} & \textbf{Stdev} & \textbf{Min} & \textbf{10pc} & \textbf{25pc} & \textbf{50pc} & \textbf{75pc} & \textbf{90pc} & \textbf{Max} \\

\hline
\multicolumn{10}{c}{\textbf{Intrinsic metrics}} \\
\hline
NoP  & 9.05   & 22.25  & 1   & 1    & 2    & 4      & 10   & 18   & 509   \\
NoS  & 1.19   & 3.09   & 0   & 0    & 0    & 1      & 1    & 3    & 83    \\
NoA  & 0.39   & 1.98   & 0   & 0    & 0    & 0      & 0    & 1    & 48    \\
NoT  & 1.50   & 3.29   & 0   & 1    & 1    & 1      & 1    & 2    & 84    \\
NoAt & 4.74   & 8.00   & 0   & 1    & 1    & 2      & 5    & 10   & 250   \\
NoCh & 267.50 & 600.44 & 8   & 57   & 84   & 153    & 267  & 489  & 22323 \\

\hline
\multicolumn{10}{c}{\textbf{Relative metrics}} \\
\hline
NCP  & 4.87   & 17.55  & 0   & 0    & 0    & 1      & 5    & 12   & 509   \\
NCS  & 0.59   & 1.96   & 0   & 0    & 0    & 0      & 1    & 2    & 82    \\
NCA  & 0.20   & 1.08   & 0   & 0    & 0    & 0      & 0    & 1    & 48    \\
NCT  & 0.85   & 2.05   & 0   & 0    & 0    & 1      & 1    & 2    & 83    \\
RED  & 10.74  & 25.93  & 0   & 0    & 2    & 4      & 12   & 24   & 1020  \\
JI   & 0.46   & 0.39   & 0   & 0    & 0    & 0.5    & 0.83 & 1    & 1     \\

\hline

\end{tabular}
}}
\caption{Average, standard deviation, range and some percentiles for query features on the SQLShare dataset 
}
\label{table:trends-query}
\end{center}
\end{table}


A first remark is that many queries have a high number of projections. Indeed, 38 queries (out of 10,668) project more than 100 expressions, while more than 21\% project more than 10 expressions. Many outliers are due to * wild card applied to large tables. 
The number of common projections, and the number of attributes 
(influenced by the number of projections) are also high. Furthermore, the number of attributes is close to the number of projections, which illustrates that predominant operations are projections. However, notice that as * wild card is not considered when counting attributes, the number of projections is higher than the number of attributes in many cases.
The number of the other query fragments is less impressive. Less than 1\% of queries exceed 10 selections, 10 aggregations or 10 tables. 
Average and standard deviation confirm this disproportion, while median values show that most queries have few projections, selections, aggregations, 
tables and attributes.

The number of characters provides a preliminary idea of query complexity, which ranges from few characters (as in query "SELECT 1") to tens of thousands characters for a small number of huge queries. Median value is 153 characters.

Focusing on longer queries (with more query fragments), at 90-percentile, queries have 18 
projections, 3 selections, 1 aggregation, 
2 tables and 10 attributes, while at 75-percentile those values are 10, 1, 0, 1 and 5 
respectively. 
Indeed, as expected, there is a large number of short queries (having less fragments):  82\% of queries have no aggregations and 44\% have no selections, while 20\% have a unique projection and 78\% have a unique table. Interestingly, 6\% of queries have no table in the FROM clause. An example of such queries is "SELECT 1+2". 

Concerning common fragments between contiguous queries, almost half of the queries have 1 common projection, and 
1 common table 
but no common selections nor aggregations, while there is more sharing at 75-percentile. 
The remaining two features, Relative Edit Distance and Jaccard Index, inform  about the combination of such common fragments. Specifically, half of the queries differ in 4 or less fragments (RED=4) and have at least 50\% of its fragments in common (JI=0.5), w.r.t. previous query. Furthermore, only 26\% of queries have nothing in common with the previous query in their sessions (JI=0).


\subsection{Comparison with the Open and Enterprise datasets}
\label{sec:extr-ground-truth}

For these two datasets, 
the feature extraction was made as in \cite{DBLP:journals/is/DjedainiDLMPV19}, enriched with estimation 
of the features pertaining to tables (NoT and NCT) and attributes (NoAt
). In addition, we included join conditions (in NoS and NCS), not considered in \cite{DBLP:journals/is/DjedainiDLMPV19} nor in our previous study \cite{DBLP:conf/dolap/PeraltaVRM19}.

Table \ref{table:datasets} compares both 
datasets in terms of number of sessions, number of explorations (the ground truth), number of queries and summarizes features extraction. The last column allows the direct comparison SQLShare values; of course without lines concerning a ground truth.

A first remark concerns the length of sessions. The Open dataset contains long sessions concerning few explorations while the Enterprise dataset contains shorter sessions concerning more explorations. 
Sessions length is actually dependent on the GUI used ; while third party OLAP tools, like Saiku, log a new query for each user action (including intermediate drag-and-drops), the SAP prototype only logs final queries. 
Regarding features, queries in both 
datasets concern a quite small number of projections, selections, aggregations, 
tables and attributes.

In addition, note that in terms of queries per session, the SQLShare dataset is similar to the Enterprise one, while in terms of features, it has bigger gaps with the other datasets. Specifically, SQLShare queries, in average, are richer in terms of projections (with high variations among queries), but contains less aggregations, selections and tables, being intermediate in terms of attributes. 
Regarding relative features, except for the number of common projections, 
most features show that queries are less similar than in the other datasets. 
Relative edit distance (RED) and Jaccard index (JI) illustrate that queries are more similar in the Enterprise dataset.


\begin{table}[]
\begin{center}
\begin{tabular}{lcc|c}
                                  & \textbf{Open}            & \textbf{Enterprise}     & \textbf{SQLShare}          \\
Nb of sessions                    & 16              & 24             & 451              \\
Nb of explorations  & 28              & 104            &               \\
Nb of queries                     & 941             & 525            & 10,668            \\
Avg queries per session               & 58         & 21         & 24            \\
Avg queries per explor.               & 34         & 5         &            \\
Avg explor. per session          & 2     & 4    &              \\
Avg and range of NoP              & 3.62 {[}1,7{]}  & 2.18 {[}0,6{]} & 9.05 {[}1,509{]} \\
Avg and range of NoS              & 3.61 {[}0,26{]} & 1.79 {[}0,5{]} & 1.19 {[}1,83{]} \\
Avg and range of NoA              & 1.34 {[}1,4{]}  & 1.14 {[}0,5{]} & 0.39 {[}0,48{]} \\
Avg and range of NoT              & 3.28 {[}1,7{]}  & 2.03 {[}1,4{]} & 1.50 {[}0,84{]} \\
Avg and range of NoAt             & 8.18 {[}1,19{]} & 4.24 {[}0,10{]}& 4.74 {[}0,250{]} \\
Avg and range of NCP              & 3.16 {[}0,7{]}  & 1.34 {[}0,4{]} & 4.87 {[}0,509{]} \\
Avg and range of NCS              & 3.12 {[}0,25{]} & 1.03 {[}0,5{]} & 0.59 {[}0,82{]} \\
Avg and range of NCA              & 1.17 {[}0,4{]}  & 0.77 {[}0,3{]} & 0.20 {[}0,48{]} \\
Avg and range of NCT              & 2.97 {[}0,7{]}  & 1.46 {[}0,4{]} & 0.85 {[}0,83{]} \\
Avg and range of RED              & 3.85 {[}0,19{]} & 2.09 {[}0,25{]}& 10.74{[}0,1020{]} \\
Avg and range of JI               & 0.57 {[}0,1{]}  & 0.79 {[}0,1{]} & 0.46 {[}0,1{]} \\
\end{tabular}
\caption{Characteristics of Open, Enterprise and SQLShare 
datasets}
\label{table:datasets}
\end{center}
\end{table}



\begin{figure}
\begin{center}
\includegraphics[width=6cm]{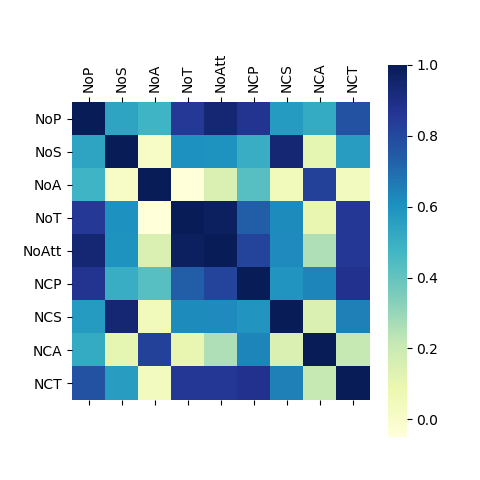}
\includegraphics[width=6cm]{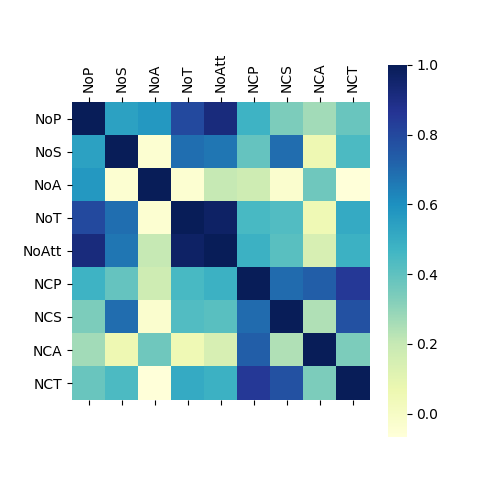}

\includegraphics[width=6cm]{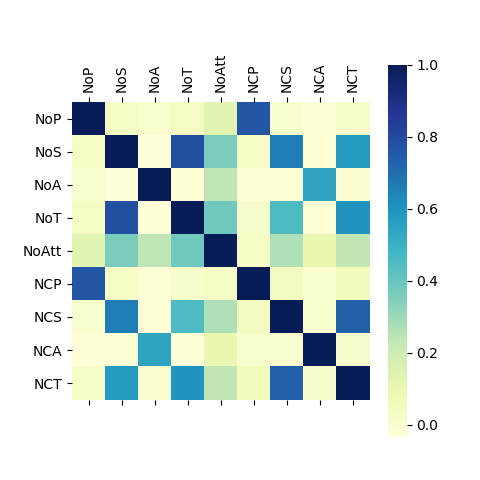}
\includegraphics[width=6cm]{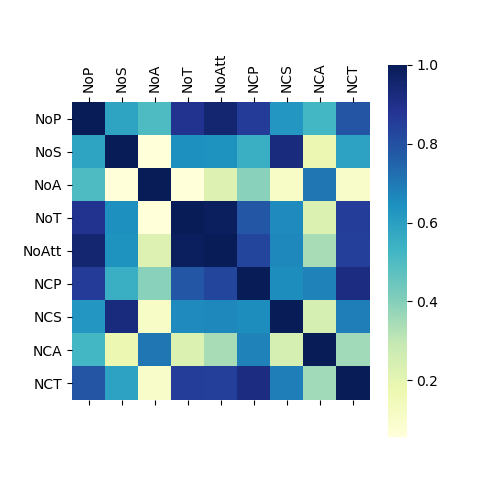}
\end{center}
\caption{Feature correlation in  datasets
Open (up left), Enterprise (up right),
SQLShare (bottom left) and Concatenate (bottom right).}
\label{fig:correlation-sqlshare}
\end{figure}

In order to complete the analysis of query features in the three datasets, we study their pairwise correlation. We exclude RED and JI features (which are aggregated from the other query parts) and NoCh (which is merely informative). 
Correlation results are shown in Figure \ref{fig:correlation-sqlshare}.


First of all, note that correlations are stronger in the Open and Enterprise datasets than in the SQLShare dataset. Specifically, the stronger correlations are those of NoAt with NoP and NoT. This is partially explained by the type of queries (OLAP-like)
and the underlying star-like database schemes. 
Noticeably, the attributes on the query text are mainly the projected attributes and the ones used for joins (and more joins are necessary when more attributes are projected). 
These attributes are less correlated in the SQLShare dataset, where both database schemes and queries are free-style. 

%
This strong correlation of NoAt with NoP and NoT, happening in the datasets used for learning (as will be explained in Section \ref{sec:sessions}) and not happening in SQLShare, reinforce the decision of not including NoAt as a core feature.

Another interesting correlation concerns NoS and NoT, being strong in all datasets, but particularly in the SQLShare dataset. Indeed, the most tables are used, the more selections are necessary for joining them. 
As expected, there are other correlations among 
intrinsic
and relative features, for example, between NoP and NCP. They happen in all datasets.

The last dataset in Figure \ref{fig:correlation-sqlshare}, called Concatenate, is the union of Open and Enterprise datasets, which is used as training set in our proposal (see Section \ref{sec:sessions}). The correlation among its attributes is very close to the one among attributes in the Open dataset.

Next section
discusses how to take into account these features for fragmenting sessions. 


\section{Session segmentation}
\label{sec:sessions}

Subsection \ref{sec:extr-sql} 
presented various statistics about queries in the SQLShare workload. A preliminary session segmentation (contiguous queries of a same user) resulted in some extremely long sessions (maximum of 937 queries) with 29\% of queries having nothing in common with their immediate predecessor.
In this section, we explore how to segment sessions in a smarter way.

Session segmentation has been previously studied for the SDSS 
workload \cite{skyserver-traffic-report-the-first-five-years}. 
In their study, 
the authors consider that a new session starts after 30 minutes of think-time (time spent between two queries).
A similar problem was largely studied for the segmentation of web traces (see for example \cite{Wong06characterizationand}) proposing the same 30-minutes cutoff. Search engine providers, like MSN and Google, use similar heuristics.  
Contrarily to those works, the published SQLShare workload does not include query timestamps.  We need to explore other heuristics for session segmentation.

In this section we explore three alternative methods for session segmentation, based respectively in the similarity between contiguous queries (Subsection \ref{sec:sim-segm}), in model reuse and transfer learning (Subsection \ref{sec:trans-segm}) and weak supervision (Subsection \ref{sec:weak-segm}).



\subsection{Similarity-based session segmentation}
\label{sec:sim-segm}

Intuitively, our idea is to compare contiguous queries in a session and segment when queries are dissimilar enough. 
Based on query features described in the previous section, we investigate 5 similarity indexes:

\paragraph{Edit Index}
It is based on the Relative Edit Distance (RED) query feature. For normalizing, RED is translated to the [0,1] interval, considering similarity is 0 after a given number of operations (arbitrarily set to 10).
\begin{equation}
\begin{aligned}
EditIndex(q_k) = max \{0, 1 - \frac{RED(q_k,q_{k-1})}{10}\}
\end{aligned}
\end{equation}

\paragraph{Jaccard Index}
It is the Jaccard Index defined in Equation \ref{eq:JaccardIndex}, which is normalized by definition.

\paragraph{Cosine Index}
It is calculated as the Cosine of vectors consisting in 
8 
query features, namely, NoP, NoS, NoA, NoT, 
NCP, NCS, NCA, and 
NCT. 
Let $x = \tuple{x_1,\ldots, x_8}$ and $y = \tuple{y_1,\ldots,y_8}$ 
be the vectors for queries $q_k$ and $q_{k-1}$ respectively.
\begin{equation}
\begin{aligned}
CosIndex(q_k,q_{k-1}) = \frac{\sum x_i . y_i}
{\sqrt{\sum x_i^2 . \sum y_i^2}}
\end{aligned}
\end{equation}

\paragraph{Common Fragments Index}
It is calculated as the number of common fragments normalized to the [0,1] interval and considering similarity is 1 when there are more than 10 common fragments (arbitrarily set). 
\begin{equation}
\begin{aligned}
CFIndex(q_k,q_{k-1}) = min \{1,  
\frac{NCF}{10}\}
\end{aligned}
\end{equation}

where 
$NCF=NCP(q_k,q_{k-1}) + NCS(q_k,q_{k-1}) + NCA(q_k,q_{k-1}) \\ 
+ NCT(q_k,q_{k-1})$.

\paragraph{Common Tables Index}
It is calculated as the number of common tables. We wanted this index to be relative to the user session ;
this is why normalization here is specifically achieved in relative terms, by dividing 
by the highest number of tables in the session.  
\begin{equation}
\begin{aligned}
CTIndex(q_k,q_{k-1}) = \frac{NCT(q_k,q_{k-1})}
{max \{NoT(q) | q \in session(q_k) \}}
\end{aligned}
\end{equation}

Note that these indexes calculate complementary aspects of query similarity 
and are normalized in different ways. Our intention is to capture different points of view and therefore to deal with different situations. 
Edit Index and Common Fragment Index count differences (resp., common fragments) as absolute values (normalized with a given threshold). Jaccard Index is a compromise of the previous ones, computing the ratio of common fragments. Cosine Index is computed using features (the value of the metrics) instead of comparing sets of fragments; it captures the variability in query complexity. And finally, Common Table Index responds to the intuition that common tables have more impact than the other common fragments, and it is normalized with respect to the number of tables used in the user session.

\begin{figure}
\begin{center}
\includegraphics[width=12cm]{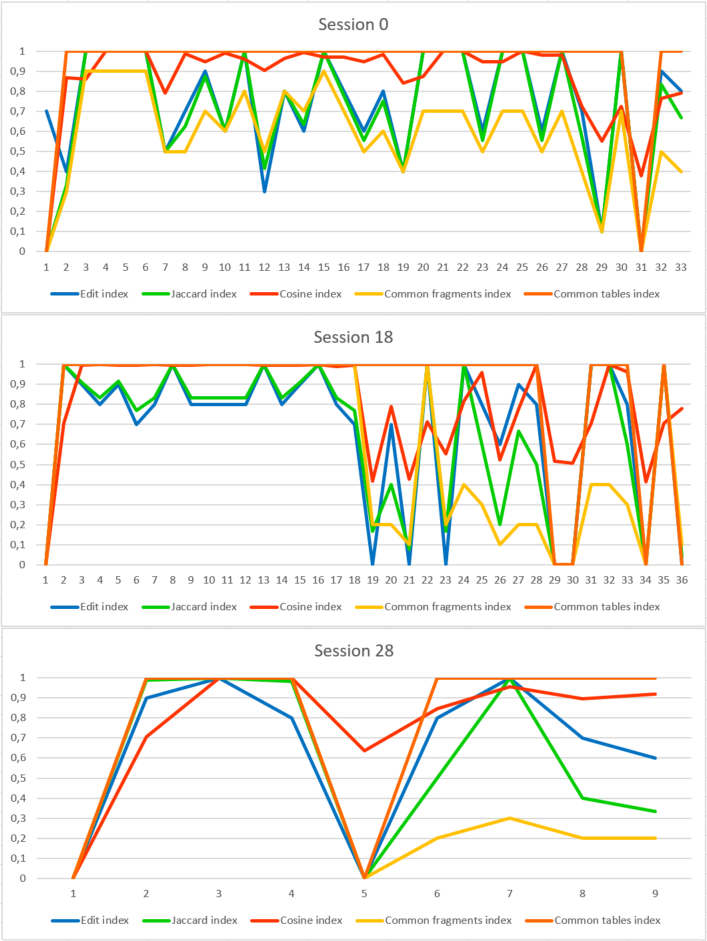}
\end{center}
\caption{Comparison of similarity indexes for 3 sessions. 
}
\label{fig:query-similarity}
\end{figure}

As an example, Figure \ref{fig:query-similarity} depicts the similarity indexes for 3  sessions of different sizes.
Looking at Session 28, the shorter one, it seems quite clear that the session may be split in two parts, by cutting between queries 4 and 5. All similarity indexes agreed.
Things are less evident for Session 0. One split seems evident (at query 31), but some others may be discussed (e.g. at queries 
29 and 12). Decision to split the session will depend on what similarity thresholds to use for the indexes.
Finally, Session 18 presents a first part, with a focused analysis, via similar queries, and a second part, more exploratory, with varied queries. Even if indexes do not always agree, their majority seems to indicate a tendency.  
%

In order to tune the similarity thresholds,
we observed the distribution of values for each similarity index (see Table \ref{table:frag-thresholds}). Many zeros indicated that a lot of queries have nothing in common with previous ones. Based on this, our intuition is to set thresholds at values around  30-percentile for each index. 
In Section \ref{sec:tests}, we experimentally tune thresholds, based both on this observation and on experiments on datasets where there is a ground truth.

In practice, our approach can be summarized as follows: For each pair of consecutive queries: (i) compute query similarity according to the proposed similarity indexes, (ii) compare the obtained similarity values with their respective thresholds, obtaining a set of votes for 
``CONTINUE" (do not segment) or ``SEGMENT" (segment and start a new exploration).
The decision (to keep consecutive queries together, or to segment) is taken by majority.


Finally, note that we propose a preliminary set of query features and similarity indexes, but the approach can easily be extended with other features and other similarity indexes.
%


\begin{table}[!ht]
\begin{center}
{\small{ 
\begin{tabular}{lccccc}
       & \textbf{Edit} & \textbf{Jaccard} & \textbf{Cosine} & \textbf{Common frag-}  & \textbf{Common}   \\
       & \textbf{index} & \textbf{index} & \textbf{index} & \textbf{ments index} & \textbf{tables index} \\
Min    & 0,00       & 0,00          & 0,05         & 0,00                   & 0,00                \\
10pc   & 0,00       & 0,00          & 0,68         & 0,00                   & 0,00                \\
20pc   & 0,00       & 0,00          & 0,72         & 0,00                   & 0,00                \\
30pc   & 0,00       & 0,10          & 0,81         & 0,10                   & 0,00                \\
40pc   & 0,40       & 0,29          & 0,89         & 0,20                   & 0,05                \\
50pc   & 0,60       & 0,50          & 0,95         & 0,30                   & 0,20                \\
60pc   & 0,80       & 0,67          & 0,99         & 0,50                   & 0,50                \\
70pc   & 0,80       & 0,80          & 1,00         & 0,60                   & 0,50                \\
80pc   & 0,90       & 0,91          & 1,00         & 0,90                   & 1,00                \\
90pc   & 1,00       & 1,00          & 1,00         & 1,00                   & 1,00                \\
Max    & 1,00       & 1,00          & 1,00         & 1,00                   & 1,00               

\end{tabular}
}}
\caption{Percentiles in distribution of indexes values.}
\label{table:frag-thresholds}
\end{center}
\end{table}

\subsection{Transfer learning based session segmentation}
\label{sec:trans-segm}

Our second method 
for segmenting the SQLShare workload is based 
on transfer learning, that consists of using supervised learning to tune a model over a labelled dataset and use this model over a dataset for which no ground truth is available. We first introduce the basics of transfer learning, and then describe our approach in details. 


\subsubsection{Transfer learning}\label{sec:transfer}

Classical supervised machine learning supposes large collections of previously collected labeled training data, to build effective predictive models. When labeled data is scarce, semi-supervised approaches may be used to build classifiers over large amount of unlabeled data and a small amount of labeled data. Still, such approaches assume that the distributions of the labeled and unlabeled data are the same.
Transfer learning, however, aims to extract the knowledge from one or more source tasks and applies the knowledge to a target task, while allowing the domains, tasks, and distributions used in training and testing to be different.  Transfer learning situations differ in what, how and when to transfer \cite{DBLP:journals/tkde/PanY10}.

In our context, having no ground truth for the SQLShare dataset, but having ground truth for other datasets, 
and considering the difference in feature correlation between SQLShare and the other datasets (see Figure \ref{fig:correlation-sqlshare}), 
allows to model session  segmentation as a classification task, and use transfer learning.  
Precisely, we will consider learning a classifier over ground truth datasets as a source task, and learning a classifier over SQLShare as the target task.
According to the typology introduced in \cite{DBLP:journals/tkde/PanY10}, this is a case of transductive transfer learning setting, where the source and target tasks are the same, while the source and target domains are different, but the feature spaces between domains are the same.
In that case, learning a model that can generalize to the target dataset demands to remove the sample selection bias due to the fact that source data and target data are drawn from different distributions. This can be achieved by reweighting the source data after having estimated the probability of appearance of each sample of the source dataset in both the source and the target dataset, 
which can be done for instance with density ratio estimation \cite{DBLP:conf/nips/SugiyamaNKBK07,DBLP:conf/nips/HuangSGBS06}.




\subsubsection{Binary classification with linear SVM}

We formalize the problem of workload segmentation as a supervised classification task, in the spirit of what we did in one of our earlier work \cite{DBLP:journals/is/DjedainiDLMPV19} 
and using datasets with ground truth that we used in this previous work.
We use a description of each query by a set of features, 
using the features that are the most correlated to the ground truth.
Our objective is to learn a linear combination of the features that separates queries starting an exploration 
from queries continuing  an exploration. 
To this end, each vectorized query of the ground truth dataset
is associated with a binary label representing the ground truth of this dataset: label SEGMENT is associated with queries starting an exploration 
(i.e. a segmentation  to be found), 
while label CONTINUE is associated with queries continuing an exploration. 

To learn our model,  we trained a  binary classifier over the 
dataset, removing sample selection bias by reweighting samples using kernel-mean matching (KMM) \cite{DBLP:conf/nips/HuangSGBS06}.
We chose a  linear SVM classifier since this  proved effective in  \cite{DBLP:journals/is/DjedainiDLMPV19}.
The model is learned 
using 10-fold cross validation to choose its best hyperparameter via randomized search. 

Since we believe that the dataset is likely to be heavily unbalanced towards the   CONTINUE label, we tested various balancing strategies while training the model, aiming at improving classification accuracy.
We compared several methods on the basis of their respective accuracy and F1-measure, over a 10-fold cross-validation: either random undersampling of majority class, or oversampling of minority class. In the last case, several heuristics have been tested: random oversampling, 3 variants of SMOTE (with different approaches to sample borderline points between classes) or ADASYN \cite{DBLP:journals/sigkdd/BatistaPM04}.

Once the best hyperparameter is obtained, the model is eventually trained over the full reweighted Concatenate dataset, to be applied over the target SQLShare dataset.

\subsection{Weak labelling and generative model}
\label{sec:weak-segm}

Instead of directly learning a transferable model from a labelled dataset, our  third approach uses a generative model to predict the labels of the unlabelled dataset. 
To this end, we resort to weak supervision, a labeling technique consisting of using noisier or heuristic sources of labels to avoid hand-labeling data. 
In our case, we wrote a set of (potentially contradictory) labeling functions, apply them to our labeled sources, using a generative model to assign the most probable label to each data, and choose the subset of functions that maximizes accuracy and F1-measure w.r.t. the ground truth.

We use Snorkel \cite{DBLP:journals/pvldb/RatnerBEFWR17}, a weak supervision system that (1) lets users write labeling functions (LFs), (2)  applies the LFs over unlabeled data and learns a generative model to combine the LFs’ outputs into probabilistic labels, and eventually  (3) allows to use these labels to train a discriminative classification model.

Snorkel is intended to work over unstructured data. Labeling functions take as input a Candidate object, representing a data point to be
classified. Each Candidate is a tuple of Context objects, which are part of a hierarchy representing the local context of the Candidate \cite{DBLP:journals/pvldb/RatnerBEFWR17}.
Typically, a candidate is a pair of named entities and the context is a sentence in which they both appear, this sentence itself being part of a document, the set of documents being the dataset to be labelled.
We adapted to Snorkel's data model by considering each session of the labelled dataset as a context, and each pair of consecutive queries in a session as a candidate. 
We wrote simple labelling functions using the metrics and indexes extracted from the datasets.
To maximize agreement between labelling functions, we grouped them and select the best subset of each group in terms of F-measure, when trained over the labelled dataset.
We then merge the best subgroups and repeat this process until  the score no longer improves.

We give below a brief description of our labelling functions.

\subsubsection{Labeling functions}

We implemented 21
labelling functions, each one using one of the relative metrics or indexes extracted from the dataset. 
Considering that SQL workloads can be very different, our objective was to define functions that capture, through simple heuristics, intuitive properties of pairs of SQL queries, and to remain independent from the dataset. 
As with the previous approach, we use a binary 
labelling scheme: 
label CONTINUE indicates 
that both queries of the pair should remain in the same exploration, 
label SEGMENT indicates that the pair should be split, i.e., the second query  starts a new exploration. 

Our first group of functions consists of one function per index (edit index, etc.), all being based on the same algorithm: if the index is greater than 0 then the pair is assigned label CONTINUE, otherwise label SEGMENT is assigned.

Our second group of functions implements a precision and a recall indicator for each of the 4 relative metrics (NCP, NCS, NCA, NCT), resulting in 8 
functions. 
For such a relative metric, say NCP,   recall (resp. precision) is computed as $\frac{NCP}{NP_f}$ (resp. $\frac{NCP}{NP_s}$) where $NP_f$ (resp. $NP_s$) is that of the first (resp. second)  query of the pair. 
All labelling functions are then based on the same algorithm: if  recall (resp. precision) equals 1 then the pair is assigned label CONTINUE,  else if it equals 0, then label is SEGMENT. Otherwise the function does not assign any label.

Our third and last group is a second implementation of precision and recall for all 4 relative metrics (another 8  
functions), favoring the attribution of the CONTINUE label, as follows: if recall (resp. precision) is not 0 then label is CONTINUE, otherwise it is SEGMENT.

\section{Experiments}
\label{sec:tests}

In this section we report the results of the experiments conducted 
to validate our proposal for session segmentation. 
Experiments with the three methods are reported, namely, similarity-based (Subsection \ref{sec:results-sim}), transfer learning-based (Subsection \ref{sec:results-trans}) and weak labelling-based (Subsection \ref{sec:results-weak}). We start by presenting our protocol 
and our baselines. 

\subsection{Protocol and baselines}

Our work aims at finding the best way of segmenting a SQL workload, upon which little is known (no timestamps, no ground truth, no database instance), into meaningful, coherent explorations. As explained in the previous section, we will test  three different segmentation methods and use datasets with ground truth to compare them. 
In what follows, we consider three datasets with ground truth: Open, Enterprise, and  the concatenation of the Open and Enterprise datasets, resulting in 40 sessions and 1466 queries, and dubbed the Concatenate dataset.
We will apply all methods on these three datasets and on the SQLShare workload, and eventually we will compute the agreement between them.

The input of our methods is a CSV file per workload (SQLShare, Open, Enterprise and Concatenate), each line describing a query by means of: query id, session id, features (extracted as explained in Section \ref{sec:extraction}), indexes (calculated as explained in Subsection \ref{sec:sim-segm}) and ground truth when available (labels SEGMENT and CONTINUE). The output of each method is an additional column in each file, indicating the segmentation (labels SEGMENT and CONTINUE).
In experiments with ground truth, both columns (ground truth and segmentation) are compared in order to evaluate the effectiveness of each method. We compute four classical quality metrics, defined as follows:

\begin{itemize}

\item \textit{Accuracy} measures the ratio of queries having the same label.
\begin{equation}
    Accuracy = \frac{nb(S,S) + nb(C,C)}
    {nb(S,S) + nb(S,C) + nb(C,S) + nb(C,C)}
\end{equation}

\item \textit{Precision} measures the ratio of queries coinciding in SEGMENT label among the queries labeled SEGMENT in the obtained segmentation. 
\begin{equation}
    Precision = \frac{nb(S,S)}
    {nb(S,S) + nb(C,S)}
\end{equation}

\item \textit{Recall} measures the ratio of queries coinciding in SEGMENT label among the ones having SEGMENT label in the ground truth. 
\begin{equation}
    Recall = \frac{nb(S,S)}
    {nb(S,S) + nb(S,C)}
\end{equation}

\item \textit{F-measure} computes the harmonic average of precision and recall.
\begin{equation}
    Fmeasure = \frac{2 * Precision * Recall}
    {Precision + Recall}
\end{equation}

\end{itemize}

where $nb(i,j)$ indicates the number of queries having label $i$ in the ground truth and label $j$ in the obtained segmentation, $i$ and $j$ being either $S$ (for SEGMENT) or $C$ (for CONTINUE). 


Noticeably, we changed the evaluation protocol compared to our previous study \cite{DBLP:conf/dolap/PeraltaVRM19}. In \cite{DBLP:conf/dolap/PeraltaVRM19}, each session was segmented independently, while our current methods process all sessions together (a whole workload). This means that our methods must also find the starts of sessions, not only the cuts inside sessions. 
%
%
Reasons are twofold: First, we can train supervised and weakly-supervised methods with bigger inputs. Second, we increased the number of SEGMENT labels to find (which makes a real difference for the Open dataset, where we pass from 12 to 28 cuts to find). We expect this change in the protocol to improve results, specially in terms of recall.

Our baseline is a naive unsupervised  clustering with automatic detection of the number of clusters using knee detection \cite{DBLP:conf/icdcsw/SatopaaAIR11,DBLP:conf/ictai/SalvadorC04}. We used a hierarchical clustering with single-linkage and Euclidean distance over queries described by the 5 indexes. On the Open, Enterprise and Concatenate dataset, results are as in Table \ref{table:baseline}. 
Note that we could have chosen as baseline 
the prediction of the majority class (i.e., never predicting SEGMENT), which would result
in accuracy being 97\% for Open, 
82\% for Enterprise and 91\% for Concatenate.
However, such baseline would have 0 as F-measure,
since recall will be 0. We will then simply use prediction of the majority class as a baseline for accuracy.

Our methods are implemented in Python, and code and data are available from  Github\footnote{\url{https://github.com/patrickmarcel/SQLWL-segmentation}}.

\begin{table}[]
\begin{center}
\begin{tabular}{l|ccc}
          & \textbf{Open} & \textbf{Enterprise}  & \textbf{Concatenate} \\
Accuracy  & 0.36 & 0.67         & 0.63             \\
Precision & 0.04    & 0.32         & 0.18             \\
Recall    & 1 & 0.69         & 0.96            \\
F-measure & 0.08 & 0.43         & 0.31             \\
\end{tabular}
\caption{Baseline results 
}
\label{table:baseline}
\end{center}
\end{table}



\subsection{Results on similarity-based session segmentation}
\label{sec:results-sim}

We implemented our first heuristic for session segmentation based on the 5 similarity indexes with voting strategy.

We first experiment with the Open and Enterprise dataset, measuring the effectiveness of our approach by comparing to the ground truth. As the Open dataset also contains timestamps, we  compare to the 30-minutes cutoff heuristic used in the literature.

\subsubsection{Experiments with ground truth} 

\paragraph{Threshold setting}
We tested different thresholds for voting. 
We proceeded as follows: we calculated the distribution of values for each index and we used as threshold the value at k-percentile, with k varying between 0 and 30. 

The thresholds that provided better results were those at 0- and 14-percentile 
for the Open and Enterprise 
datasets respectively. These thresholds reflect the relationship between the number of explorations to find and the number of queries, as well as the similarity among consecutive queries. Indeed, the Open dataset contains many queries and few explorations (i.e., a few segments to find); small thresholds are best adapted. Conversely, the Enterprise dataset needs to be more segmented as the average number of queries per exploration is low; higher thresholds do better.

We remark that more precise thresholds could be learned with supervised machine learning techniques (e.g. classification). 
We intentionally avoid this computation in this first method, purely unsupervised, because in real applications (like with SQLShare) we do not have any ground truth for learning.
An expert providing the ratio of queries per exploration (either intuitively or via preliminary tests) is more realistic. 
Besides, the use of supervised learning is analysed in 
the next subsections.


In the remaining tests, we use values at 0- and 14-percentile 
as thresholds for the Open and Enterprise 
datasets, respectively.

\paragraph{Segmentation quality}
For each dataset, we compared the obtained segmentation to the ground truth, measuring segmentation quality in 
terms of accuracy, precision, recall and F-measure.
%
%
We report the results in Table \ref{table:segm-quality}.

As expected, results are very good in terms of accuracy, mainly explained because classes are unbalanced (the number of CONTINUE labels is higher than the number of SEGMENT ones)
and quite good in terms of f-measure; better for the Open dataset than for the Enterprise one.
%
%
Note that results are different from those reported in \cite{DBLP:conf/dolap/PeraltaVRM19} because the evaluation protocol changed, as previously explained.

\begin{table}[]
\begin{center}
\begin{tabular}{l|ccc|c}
          & \textbf{Open} & \textbf{Enterprise} & \textbf{Concatenate} & \textbf{Open} (timestamp) \\
Accuracy  & 0.99 & 0.93  & 0.97         & 0.99             \\
Precision & 1    & 0.85  & 0.88         & 1             \\
Recall    & 0.75 & 0.72  & 0.73         & 0.64             \\
F1-measure & 0.86 & 0.78 & 0.80         & 0.78             \\
\end{tabular}
\caption{Segmentation results for our approach on the 3 datasets and the timestamp-based approach (rightmost column)
}
\label{table:segm-quality}
\end{center}
\end{table}

\paragraph{Comparison to timestamp-based approach}
In order to compare our approach to the one used in the literature, we implemented a second heuristic that segments users sessions when there is a 30-minutes delay between queries. The Open dataset, the only one containing timestamps, was used for this test. 
Results are reported in Table \ref{table:segm-quality}, the right-most column corresponding to the timestamp-based approach. They are comparable in terms of accuracy and  
lower in terms of f-measure. Note that 1 for precision means that all 
cuts
found are also breaks in the ground truth. 
In other words, there are no big delays inside explorations, which makes sense. However, the timestamp-based approach fails to detect 36\% of the breaks (when the user changes its topic of study in a briefer delay).

\paragraph{Analysis of similarity indexes}
Finally, we investigated the quality of the 5 proposed similarity indexes, by studying the correlation of their vote (when index value is lower than the corresponding threshold) with respect to the breaks in the ground truth. Results are presented in Table \ref{table:correlation} (left). 

Jaccard and CF indexes are the more correlated in the Enterprise dataset. Both of them are highly correlated in the Open dataset, CT index being even better. Edit and Cosinus indexes are less correlated in one of the datasets. 
Interestingly, the most influencing indexes, the ones more correlated with the final vote (as shown in Table \ref{table:correlation} (right)) are also Jaccard, CF and CT indexes.


\begin{table}[]
\begin{center}
\begin{tabular}{lcc|cc}
              & \multicolumn{2}{c|}{\textbf{Ground truth}} &  \multicolumn{2}{c}{\textbf{Final vote}}\\
              & \textbf{Open} & \textbf{Enterprise} & \textbf{Open} & \textbf{Enterprise}  \\
              \hline
Edit index    & 0.30 & 0.63 & 0.20 & 0.69\\
Jaccard index & 0.86 & 0.72 & 1.00 & 0.97 \\
Cos index     & 0.75 & 0.36 & 0.87 & 0.51\\
CF index      & 0.86 & 0.69 & 1.00 & 0.88\\
CT index      & 0.90 & 0.50  & 0.95 & 0.65 
\end{tabular}
\caption{Correlation between votes of similarity indexes and ground truth (left) / final vote (right) for both 
datasets.}
\label{table:correlation}
\end{center}
\end{table}

\subsubsection{Experiments with SQLShare}



As in experiments with ground truth, we use our 
heuristic based in the 5 similarity indexes, tuned with simple thresholds
, taking the decision of segmenting or not using majority vote. 

According to the findings on Open and Enterprise dataset, we set thresholds for the SQLShare dataset based
on statistics, like number of queries per session (see Table \ref{table:datasets}) and the dataset description in \cite{DBLP:conf/sigmod/JainMHHL16}. The proposed thresholds are the 
values at 30-percentile, which are coherent with the preliminary analysis shown in Table \ref{table:frag-thresholds}. 

This simple heuristic allowed to split the initial 451 sessions in 2,851 
explorations. 
In the absence of ground-truth, we present in Table  \ref{table:session-stats} a comparison of average 
features before and after session segmentation.
A first remark concerns session length: extremely large sessions (maximum of 937) were split (new maximum is 98 queries). Indeed, more than half of the sessions were not fragmented and at 3rd quartile 1 session was split in 3 explorations. Some long and anarchic sessions (such as the one counting 937 queries) were split in a multitude of explorations. 
We can also highlight an increasing in the average number of common query fragments (NCP, NCS, NCA, NCT) per session. This increasing is quite regular and visible for all quartiles. 
Relative edit distance (RED) and Jaccard Index (JI) also improved, as expected.

{\footnotesize
\begin{table}[]
\begin{tabular}{l|rrrrrrr}
\multicolumn{8}{c}{\textbf{Before segmentation}}   \\
& \textbf{Avg} & \textbf{Stddev} & \textbf{Min} & \textbf{25pc} & \textbf{50pc} & \textbf{75pc} & \textbf{Max}   \\
Nb queries & 23.65 & 75.05  & 1           & 2           & 4    & 13.5 & 937    \\
Avg NCP    & 3.96  & 15.84  & 0           & 0           & 1    & 3.35 & 286.50 \\
Avg NCS    & 0.29  & 0.59   & 0           & 0           & 0    & 0.50 & 7.17   \\
Avg NCA    & 0.12  & 0.34   & 0           & 0           & 0    & 0.05 & 3.44   \\
Avg NCT    & 0.59  & 0.53   & 0           & 0           & 0.59 & 0.86 & 3.85   \\
Avg RED    & 9.20  & 15.85  & 0           & 3           & 5.33 & 9.46 & 192.38 \\
Avg JI     & 0.34  & 0.26   & 0           & 0           & 0.39 & 0.55 & 1.00   \\
\hline
\multicolumn{8}{c}{\textbf{After segmentation}} \\
& \textbf{Avg} & \textbf{Stddev} & \textbf{Min} & \textbf{25pc} & \textbf{50pc} & \textbf{75pc} & \textbf{Max}   \\
Nb queries & 3.74  & 6.16   & 1           & 1           & 1.5  & 4    & 98     \\
Avg NCP    & 6.58  & 16.91  & 0           & 1           & 3    & 7.00 & 509.00 \\
Avg NCS    & 0.70  & 3.19   & 0           & 0           & 0    & 1.00 & 82.00  \\
Avg NCA    & 0.30  & 1.59   & 0           & 0           & 0    & 0.00 & 32.00  \\
Avg NCT    & 1.24  & 4.02   & 0           & 0.77 & 1.00 & 1.00 & 83.00  \\
Avg RED    & 8.70  & 21.26  & 0           & 1.67 & 3.20 & 7.46 & 508.00 \\
Avg JI     & 0.61  & 0.28   & 0.00 & 0.42 & 0.62 & 0.83 & 1.00  

\end{tabular}
\caption{Comparison of average features per session before and after segmentation}
\label{table:session-stats}
\end{table}
}

\subsection{Results on transfer learning-based session segmentation}
\label{sec:results-trans}

In this second approach, we learn a model over our labelled dataset (source task) and transfer the model over SQLShare (target task). Our source task consists of building a SVM classifier for the Concatenate dataset.

We applied  our approach as described in Section \ref{sec:transfer}. 
We first selected, from the set of all metrics and indexes extracted from the Concatenate dataset (see Section \ref{sec:extraction} and \ref{sec:sim-segm}), the ones most correlated with the ground truth, using Pearson's correlation coefficient.
To select the most correlated features, 
we ordered the features by the absolute value of the correlation coefficient, we train the classifier iteratively adding the features while F-measure and accuracy scores increase, and stop when the scores no more improve. This resulted in keeping features 
 Edit index, Cosine index,
Common fragments index  and Jaccard index, 
whose absolute values of correlation coefficient range from  0.52 to  0.59.


As reported in Subsection \ref{sec:extr-ground-truth}, the Concatenate dataset is highly unbalanced.
Label SEGMENT appears only 124 times while label CONTINUE appears 1342 times. 
We tested different balancing strategies and different splitting 
in training and test set (from 50 to 95 percent for the training set).
Our best classifier's scores and weights are given in Table \ref{table:svm}. It is obtained using vanilla SMOTE for oversampling and  a training set of 95\%.
We remark that the classifier achieves good scores on the training sets (for hyper-parameter learning), the test set and on the complete dataset.
Unsurprisingly, the classifier weights follow the correlation of the features to the ground truth (from Jaccard the most correlated to Edit Index the less correlated). 

Trained over the complete Concatenate dataset, using KMM re-weighting, and applied over the SQLShare dataset, the classifier detected 3,437 explorations. 


{\footnotesize
\begin{table}[]
\begin{center}
\begin{tabular}{lll}
\hline
{\bf Training set}  \\
\hline
size before sampling &   1392 \\
size after sampling  &  2548 \\
Average accuracy  &  0.93 +/- 0.02\\
Average precision  &  0.91 +/- 0.01\\
Average recall  &  0.95 +/- 0.03\\
Average F-measure  &  0.93 +/- 0.02\\
\hline 
{\bf Test set} & \\
\hline 
Size& 74\\
Accuracy  &  0.94\\
Precision  &  0.6 \\
Recall &  1 \\
F-measure  &  0.75\\
\end{tabular}
\hfill
\begin{tabular}{ll}
\hline
{\bf Classifier weights}  \\
\hline
Jaccard index & -1.38 \\
Cosine index & -0.52\\
CFI & -1.0\\
Edit index & -0.50 \\
\\
\\
\hline 
{\bf Overall scores} & \\
\hline 
Accuracy  &  0.95\\
Precision  &  0.76 \\
Recall &   0.76\\
F-measure  &  0.76\\
\end{tabular}
\caption{Results of the best linear SVM classifier over Concatenate, using oversampling and transfer learning (left), classifier weights when trained over the Concatenate dataset (up right) 
and overall scores on the Concatenate dataset (bottom right)
}
\label{table:svm}
\end{center}
\end{table}
}


\subsection{Results on weak labelling-based session segmentation}
\label{sec:results-weak}

For 
this last method, the first task consists in the selection of the most appropriate subset of labelling functions. Our goal is to select the best subset,
in the sense of F-measure, 
over the Concatenate dataset.
Over the $2^{21}-1$ possible subsets, we tested 564 combinations of labelling functions, using the protocol explained in Section \ref{sec:weak-segm}.
The best subset was formed of three functions:
two functions taken in the first group (Edit index and Jaccard index) and
one function taken in the third group (the recall of projections, second version).
This subset achieves the scores displayed in Table \ref{table:labelling-scores},
which are slightly better than the one achieved with the classifier of second method.

{\footnotesize
\begin{table}[h]
\begin{center}
\begin{tabular}{ll}
Accuracy & 0.96\\
Precision & 0.76	\\
Recall & 0.76\\
F-measure & 0.76\\
\end{tabular}
\caption{Scores of the best set of labelling functions}
\label{table:labelling-scores}
\end{center}
\end{table}
}

Applied over the SQLShare dataset, this method detected 3,208 explorations. 

\subsection{Agreements}
\label{sec:results-agreements}

We now report the agreement of the three different methods applied on the Concatenate dataset and on the SQLShare workload.  The agreement is computed using the following indicators:
\begin{itemize}
    \item Pairwise Cohen's kappa, which is a  statistic measure of agreement, classically used to evaluate the degree of agreement of two raters, ranging from negative values (worse agreement than random)  to 1 (perfect agreement), with 0 indicating no agreement among the raters other than what would be expected by chance,
     \item Fleiss' kappa, that  generalizes the Cohen's Kappa to more than two raters, 
    \item Full agreement ratio, i.e., the percentage of times all methods perfectly agree,
    \item Confusion matrix: the number of each combination of predictions for the three methods.
\end{itemize}

For the Concatenate dataset, 
the overall Fleiss' kappa for the three methods is 0.91
and the full agreement ratio is 0.98.
Table \ref{table:agreement-concatenate} shows for each method its Cohen's Kappa when compared to another method, highlighting that the best agreement occurs between transfer- and weak-labelling-based methods, being two methods learning over Concatenate as ground truth. The agreement of the three methods with the ground truth is comparable and remains good. These values are frequently qualified as a strong agreement. 
Table \ref{table:confusion-concatenate} shows the  confusion matrix. 
As expected, the agreement on  CONTINUE label is the highest. Noticeably, the agreement on SEGMENT label is also very high, compared to non-agreement combinations.

{\footnotesize
\begin{table}[h]
\begin{center}
\begin{tabular}{lcccc}
& Voting & Transfer & Weak-labelling & Ground truth\\
Voting & 1 & 0.9 & 0.89 & 0.78\\
Transfer &0.9  &  1& 0.95 & 0.74\\
Weak-labelling & 0.89  &0.95 & 1 & 0.74\\
Ground truth & 0.78& 0.74 & 0.74 &1
\end{tabular}
\caption{Agreement between the three different methods and with the ground truth, on the Concatenate dataset, measured with Cohen's kappa}
\label{table:agreement-concatenate}
\end{center}
\end{table}
}

{\footnotesize
\begin{table}[h]
\begin{center}
\begin{tabular}{ccccc}
 Voting & Transfer & Weak-labelling & Percentage \\ 
0 & 0 & 0 & 0.91\\
0 & 0 & 1 & 0.004 \\
0 & 1 & 0 & 0.003 \\
0 & 1 & 1 & 0.01 \\
1 & 0 & 0 & 0\\
1 & 0 & 1 & 0 \\
1 & 1 & 0 & 0 \\
1 & 1 & 1 & 0.07 \\
\end{tabular}
\caption{Confusion matrix of the three different methods on the Concatenate dataset (0=CONTINUE, 1=SEGMENT)}
\label{table:confusion-concatenate}
\end{center}
\end{table}
}

For the SQLShare workload, the overall Fleiss' kappa for the three methods is 0.87 and the full agreement ratio is 0.91.
Table \ref{table:agreement-sqlshare} shows for each methods its Cohen's Kappa when compared to another method, and Table \ref{table:confusion-sqlshare} shows the  confusion matrix. 
Results are comparable to those for the Concatenate dataset in terms of agreement. Nevertheless, the three methods agree to segment more frequently than on the Concatenate dataset. 

{\footnotesize
\begin{table}[h]
\begin{center}
\begin{tabular}{lccc}
          & Voting & Transfer & Weak-labelling \\
Voting    & 1      & 0.84     & 0.88\\
Transfer  & 0.84   &  1       & 0.89\\
Weak-labelling & 0.88  & 0.89    & 1
\end{tabular}
\caption{Agreement between the three different methods on the SQLShare workload, measured with Cohen's kappa}
\label{table:agreement-sqlshare}
\end{center}
\end{table}
}

{\footnotesize
\begin{table}[h]
\begin{center}
\begin{tabular}{ccccc}
 Voting & Transfer & Weak-labelling & Percentage \\ 
0 & 0 & 0 & 0.66\\
0 & 0 & 1 & 0.008 \\
0 & 1 & 0 & 0.03 \\
0 & 1 & 1 & 0.03 \\
1 & 0 & 0 & 0.0001\\
1 & 0 & 1 & 0.005 \\
1 & 1 & 0 & 0.007 \\
1 & 1 & 1 & 0.26 \\
\end{tabular}
\caption{Confusion matrix of the three different methods on the SQLShare workload (0=CONTINUE, 1=SEGMENT)}
\label{table:confusion-sqlshare}
\end{center}
\end{table}
}

As a final remark, we observe that all three methods have good agreements over the test set (SQLShare). This is quite expected for Voting and  Transfer since they share most features (4 of the indexes), but a bit more surprising for Weak-labelling   since it uses Edit index (the weakest weighted in Transfer's classifier) and uses two non index-features (namely NoP and NCP
for computing the recall of projections).
Importantly, all methods agree on finding
more than 26\% of segmenting, 
consistently with our analysis of the SQLShare workload reported in Section \ref{sec:extr-sql},
as there was 26\% of queries 
having nothing in common with their immediate predecessor.
We note that of all three methods, Voting 
is the one that detects less explorations (2851 against 3437 and 3208) and achieves the best results over the ground truth dataset. 

 



\section{Conclusion}
\label{sec:conclusion}

This paper discussed the problem of segmenting sequences of SQL queries into meaningful explorations when only the query text is available, and it is not possible to rely on query timestamps.

We characterized queries as
a set of simple features and defined five  similarity indexes with respect to previous queries in the session. A simple unsupervised heuristic, based on the similarity indexes with voting strategy, allowed to split long and heterogeneous sessions into smaller explorations where queries have more connections. We investigated two additional methods, exploiting supervised and weak-supervised learning techniques. Experiments showed a strong agreement among the 3 methods; the best results, in terms of accuracy and F-measure over  datasets with ground truth, being achieved by the simple unsupervised method.

Our approach can be easily extended  with other query features and other similarity indexes. In the near future, we would like to test more features, in particular for considering each query in the context of its session (not only comparing it to its immediate predecessor) and exploiting query results.
Further similarity indexes may be deduced from such features. We also should discard preliminary hypothesis about chronological ordering of queries and deal with query similarity beyond it.

Our long term goal is to show how our segmentation approaches help improving a variety of novel log-based applications, from the  
 measurement of the quality of SQL explorations,  the detection of specific exploratory activities, the learning of user analysis behavior, the discovery of latent user intents, or the recommendation of forthcoming exploration  queries.





\bibliographystyle{abbrv}
\bibliography{biblio}

\begin{thebibliography}{10}

\bibitem{DBLP:journals/sigkdd/BatistaPM04}
G.~E. A. P.~A. Batista, R.~C. Prati, and M.~C. Monard.
\newblock A study of the behavior of several methods for balancing machine
  learning training data.
\newblock {\em {SIGKDD} Explorations}, 6(1):20--29, 2004.

\bibitem{DBLP:conf/mmm/BhattaraiWS07}
B.~D. Bhattarai, M.~Wong, and R.~Singh.
\newblock Discovering user information goals with semantic website media
  modeling.
\newblock In {\em {MMM} {(1)}}, volume 4351 of {\em Lecture Notes in Computer
  Science}, pages 364--375. Springer, 2007.

\bibitem{DBLP:conf/vldb/ChaudhuriN07}
S.~Chaudhuri and V.~R. Narasayya.
\newblock Self-tuning database systems: {A} decade of progress.
\newblock In {\em Proceedings of the 33rd International Conference on Very
  Large Data Bases, University of Vienna, Austria, September 23-27, 2007},
  pages 3--14, 2007.

\bibitem{DBLP:journals/is/DjedainiDLMPV19}
M.~Djedaini, K.~Drushku, N.~Labroche, P.~Marcel, V.~Peralta, and W.~Verdeaux.
\newblock Automatic assessment of interactive {OLAP} explorations.
\newblock {\em Inf. Syst.}, 82:148--163, 2019.

\bibitem{DBLP:conf/adbis/DjedainiLMP17}
M.~Djedaini, N.~Labroche, P.~Marcel, and V.~Peralta.
\newblock Detecting user focus in {OLAP} analyses.
\newblock In {\em Advances in Databases and Information Systems - 21st European
  Conference, {ADBIS} 2017, Nicosia, Cyprus, September 24-27, 2017,
  Proceedings}, pages 105--119, 2017.

\bibitem{DrushkuLMP19}
K.~Drushku, N.~Labroche, P.~Marcel, and V.~Peralta.
\newblock Interest-based recommendations for business intelligence users.
\newblock {\em To appear in Information Systems}, 2019.
\newblock https://doi.org/10.1016/j.is.2018.08.004.

\bibitem{DBLP:journals/tkde/EirinakiAPS14}
M.~Eirinaki, S.~Abraham, N.~Polyzotis, and N.~Shaikh.
\newblock Querie: Collaborative database exploration.
\newblock {\em {IEEE} Trans. Knowl. Data Eng.}, 26(7):1778--1790, 2014.

\bibitem{NIPS2015_5872}
M.~Feurer, A.~Klein, K.~Eggensperger, J.~Springenberg, M.~Blum, and F.~Hutter.
\newblock Efficient and robust automated machine learning.
\newblock In C.~Cortes, N.~D. Lawrence, D.~D. Lee, M.~Sugiyama, and R.~Garnett,
  editors, {\em Advances in Neural Information Processing Systems 28}, pages
  2962--2970. Curran Associates, Inc., 2015.

\bibitem{DBLP:conf/nips/HuangSGBS06}
J.~Huang, A.~J. Smola, A.~Gretton, K.~M. Borgwardt, and B.~Sch{\"{o}}lkopf.
\newblock Correcting sample selection bias by unlabeled data.
\newblock In {\em Advances in Neural Information Processing Systems 19,
  Proceedings of the Twentieth Annual Conference on Neural Information
  Processing Systems, Vancouver, British Columbia, Canada, December 4-7, 2006},
  pages 601--608, 2006.

\bibitem{DBLP:conf/sigmod/IdreosPC15}
S.~Idreos, O.~Papaemmanouil, and S.~Chaudhuri.
\newblock Overview of data exploration techniques.
\newblock In {\em Proceedings of the 2015 {ACM} {SIGMOD} International
  Conference on Management of Data, Melbourne, Victoria, Australia, May 31 -
  June 4, 2015}, pages 277--281, 2015.

\bibitem{DBLP:conf/sigmod/JainMHHL16}
S.~Jain, D.~Moritz, D.~Halperin, B.~Howe, and E.~Lazowska.
\newblock Sqlshare: Results from a multi-year sql-as-a-service experiment.
\newblock In {\em Proceedings of the 2016 International Conference on
  Management of Data, {SIGMOD} Conference 2016, San Francisco, CA, USA, June 26
  - July 01, 2016}, pages 281--293, 2016.

\bibitem{DBLP:journals/pvldb/KhoussainovaKBS11}
N.~Khoussainova, Y.~Kwon, M.~Balazinska, and D.~Suciu.
\newblock Snipsuggest: Context-aware autocompletion for {SQL}.
\newblock {\em {PVLDB}}, 4(1):22--33, 2010.

\bibitem{DBLP:journals/tkde/KulLXCKU18}
G.~Kul, D.~T.~A. Luong, T.~Xie, V.~Chandola, O.~Kennedy, and S.~J. Upadhyaya.
\newblock Similarity metrics for {SQL} query clustering.
\newblock {\em {IEEE} Trans. Knowl. Data Eng.}, 30(12):2408--2420, 2018.

\bibitem{DBLP:conf/edbt/NguyenBBGHM15}
H.~V. Nguyen, K.~B{\"{o}}hm, F.~Becker, B.~Goldman, G.~Hinkel, and
  E.~M{\"{u}}ller.
\newblock Identifying user interests within the data space - a case study with
  skyserver.
\newblock In {\em {EDBT}}, pages 641--652. OpenProceedings.org, 2015.

\bibitem{DBLP:journals/tkde/PanY10}
S.~J. Pan and Q.~Yang.
\newblock A survey on transfer learning.
\newblock {\em {IEEE} Trans. Knowl. Data Eng.}, 22(10):1345--1359, 2010.

\bibitem{DBLP:conf/dolap/PeraltaVRM19}
V.~Peralta, W.~Verdeaux, Y.~Raimont, and P.~Marcel.
\newblock Qualitative analysis of the sqlshareworkload for session
  segmentation.
\newblock In {\em Proceedings of the 21st International Workshop on Design,
  Optimization, Languages and Analytical Processing of Big Data, co-located
  with {EDBT/ICDT} Joint Conference, DOLAP@EDBT/ICDT 2019, Lisbon, Portugal,
  March 26, 2019.}, 2019.

\bibitem{DBLP:journals/pvldb/RatnerBEFWR17}
A.~Ratner, S.~H. Bach, H.~R. Ehrenberg, J.~A. Fries, S.~Wu, and C.~R{\'{e}}.
\newblock Snorkel: Rapid training data creation with weak supervision.
\newblock {\em {PVLDB}}, 11(3):269--282, 2017.

\bibitem{DBLP:conf/dawak/RomeroMAPB11}
O.~Romero, P.~Marcel, A.~Abell{\'{o}}, V.~Peralta, and L.~Bellatreche.
\newblock Describing analytical sessions using a multidimensional algebra.
\newblock In {\em Data Warehousing and Knowledge Discovery - 13th International
  Conference, DaWaK 2011, Toulouse, France, August 29-September 2,2011.
  Proceedings}, pages 224--239, 2011.

\bibitem{DBLP:conf/ictai/SalvadorC04}
S.~Salvador and P.~Chan.
\newblock Determining the number of clusters/segments in hierarchical
  clustering/segmentation algorithms.
\newblock In {\em 16th {IEEE} International Conference on Tools with Artificial
  Intelligence {(ICTAI} 2004), 15-17 November 2004, Boca Raton, FL, {USA}},
  pages 576--584, 2004.

\bibitem{DBLP:conf/icdcsw/SatopaaAIR11}
V.~Satopaa, J.~R. Albrecht, D.~E. Irwin, and B.~Raghavan.
\newblock Finding a "kneedle" in a haystack: Detecting knee points in system
  behavior.
\newblock In {\em 31st {IEEE} International Conference on Distributed Computing
  Systems Workshops {(ICDCS} 2011 Workshops), 20-24 June 2011, Minneapolis,
  Minnesota, {USA}}, pages 166--171, 2011.

\bibitem{skyserver-traffic-report-the-first-five-years}
V.~Singh, J.~Gray, A.~Thakar, A.~S. Szalay, J.~Raddick, B.~Boroski,
  S.~Lebedeva, and B.~Yanny.
\newblock Skyserver traffic report - the first five years.
\newblock Technical report, December 2006.

\bibitem{DBLP:conf/nips/SugiyamaNKBK07}
M.~Sugiyama, S.~Nakajima, H.~Kashima, P.~von B{\"{u}}nau, and M.~Kawanabe.
\newblock Direct importance estimation with model selection and its application
  to covariate shift adaptation.
\newblock In {\em Advances in Neural Information Processing Systems 20,
  Proceedings of the Twenty-First Annual Conference on Neural Information
  Processing Systems, Vancouver, British Columbia, Canada, December 3-6, 2007},
  pages 1433--1440, 2007.

\bibitem{DBLP:conf/scam/BrinkLV07}
H.~van~den Brink, R.~van~der Leek, and J.~Visser.
\newblock Quality assessment for embedded {SQL}.
\newblock In {\em {SCAM}}, pages 163--170. {IEEE} Computer Society, 2007.

\bibitem{jainComplexity}
A.~Vashistha and S.~Jain.
\newblock Measuring query complexity in sqlshare workload.
\newblock https://uwescience.github.io/sqlshare/pdfs/Jain-Vashistha.pdf.

\bibitem{DBLP:books/cu/W2016}
R.~W. White.
\newblock {\em Interactions with Search Systems}.
\newblock Cambridge University Press, 2016.

\bibitem{Wong06characterizationand}
M.~Wong, B.~Bhattarai, and R.~Singh.
\newblock Characterization and analysis of usage patterns in large multimedia
  websites.
\newblock Technical report, 2006.

\end{thebibliography}

\end{document}